\documentclass[pre,preprint,showpacs,eqsecnum]{revtex4}
\usepackage{bm}
\usepackage{graphicx}
\usepackage{amsmath,amssymb}

\begin{document}
\title{Asymptotic solutions of decoupled continuous-time random
walks with superheavy-tailed waiting time and heavy-tailed jump
length distributions}
\author{S.~I.~Denisov$^{1,3}$}
\email{denisov@sumdu.edu.ua}
\author{S.~B.~Yuste$^{2}$}
\author{Yu.~S.~Bystrik$^{1}$}
\author{H.~Kantz$^{3}$}
\author{K.~Lindenberg$^{4}$}
\affiliation{$^{1}$Sumy State University, Rimsky-Korsakov Street 2,
UA-40007 Sumy, Ukraine\\
$^{2}$Departamento de F\'{\i}sica, Universidad de Extremadura,
E-06071 Badajoz, Spain\\
$^{3}$Max-Planck-Institut f\"{u}r Physik komplexer Systeme,
N\"{o}thnitzer Stra{\ss}e 38, D-01187 Dresden, Germany\\
$^4$Department of Chemistry and Biochemistry 0340 and BioCircuits
Institute, University of California, San Diego, La Jolla, California
92093-0340, USA}


\begin{abstract}
We study the long-time behavior of decoupled continuous-time random
walks characterized by superheavy-tailed distributions of waiting
times and symmetric heavy-tailed distributions of jump lengths. Our
main quantity of interest is the limiting probability density of the
position of the walker multiplied by a scaling function of time. We
show that the probability density of the scaled walker position
converges in the long-time limit to a non-degenerate one only if the
scaling function behaves in a certain way. This function as well as
the limiting probability density are determined in explicit form.
Also, we express the limiting probability density which has heavy
tails in terms of the Fox $H$-function and find its behavior for
small and large distances.
\end{abstract}

\maketitle

\section{INTRODUCTION}
\label{Intr}

Continuous-time random walks (CTRWs), introduced by Montroll and
Weiss \cite{MW}, constitute an important class of jump processes that
are widely used to model a variety of physical, geological,
biological, economic and other phenomena. In particular, these
processes describe anomalous diffusion and transport in disordered
media (see, e.g., Refs.~\cite{AH,MK, KRS} and references therein),
seismic \cite{HS,PAG} and financial \cite{MMW,Scal} data. A
remarkable fact is that systems so different from one another can
successfully be described within the CTRW approach. This is because
two random variables that many systems have in common, the waiting
time between successive jumps and the jump length, are used to model
the CTRW. Therefore, even the decoupled CTRW, when these variables
are independent, is rather flexible.

The probability density $P(x,t)$ of the walker position $X(t)$ is the
most important characteristic of the CTRW. It satisfies the integral
master equation \cite{MBK, MRGS, Metz} which in the decoupled case
depends only on the probability density $p(\tau)$ of waiting times
and on the probability density $w(x)$ of jump lengths. Because exact
solutions of this equation are known in very few cases \cite{KZ,
Bar1,  Bar2, GPSS}, there is considerable interest in studying the
long-time behavior of $P(x,t)$ that is responsible for the transport
and diffusion properties of objects described by the CTRW model. In
this context, much attention has been paid to the probability
densities $p(\tau)$ and $w(x)$ having finite second moments and/or to
those having heavy tails. It has been established \cite{Tun,SKW, WWH}
that different combinations of these properties of the waiting time
and jump densities lead to different long-time distributions of
$X(t)$. In Ref.~\cite{Kot}, all possible distributions were expressed
in terms of the limiting distributions of the properly scaled walker
position.

In some cases the waiting-time densities are assumed to be
superheavy-tailed, i.e., such that all fractional moments of
$p(\tau)$ are infinite. In particular, this class of densities is
used to model the superslow diffusion in which the diffusion front
spreads more slowly than any positive power of time \cite{HW, DrKl,
CKS, DK1}. In general, one might expect superheavy-tailed
distributions to reflect extremely slow time-dependent phenomena such
as may occur in some relaxation and aging processes. Such
distributions are also applicable within a Langevin rather than a
CTRW description when dealing with processes that are interrupted by
an absorption event or by the transition of a particle to a
qualitatively different state \cite{DKH, DK3}. The long-time behavior
of the decoupled CTRWs characterized by these waiting-time densities
and jump densities with finite second moments is considered in
Ref.~\cite{DK2}. Here we focus on asymptotic solutions of the CTRWs
in the case when the densities $p(\tau)$ and $w(x)$ are superheavy-
and heavy-tailed, respectively.

The paper is structured as follows. In Sec.~\ref{Def}, we formulate
the main definitions and write the basic equations describing the
decoupled CTRW. A one-parameter limiting probability density of the
scaled walker position that corresponds to the superheavy-tailed
distributions of waiting times and the symmetric heavy-tailed
distributions of jump lengths is determined in Sec.~\ref{Lim}. Here,
we also find the scaling function and prove the positivity and
unimodality of the limiting probability density. In Sec.~\ref{Rep},
we express the limiting density and the corresponding cumulative
distribution function in terms of Fox $H$-functions and consider a
few particular examples. The short- and long-distance behavior of the
limiting density is studied in Sec.~\ref{Asymp}. Our main results are
summarized in Sec.~\ref{Concl}.

\section{MAIN DEFINITIONS AND BASIC EQUATIONS}
\label{Def}

The CTRW approach deals with a wide class of continuous-time jump
processes $X(t)$ represented as
\begin{equation}
    X(t) = \sum_{n=1}^{N(t)} x_{n}.
\label{X(t)}
\end{equation}
Here, $N(t) = 0,1,2,\ldots$ is the random number of jumps that a
walker has performed up to the time $t$ (if $N(t)=0$ then $X(t)=0$),
and $x_{n} \in (-\infty, \infty)$ are the independent random
variables (jump lengths) distributed with some probability density
$w(x)$. In order to specify the counting process $N(t)$, the waiting
times $\tau_{n}$, i.e., times between successive jumps, are
introduced. Like the jump lengths, the waiting times are assumed to
be independent random variables distributed with probability density
$p(\tau)$. If the variables $x_{n}$ and $\tau _{n}$ are independent
of each other as well, i.e., if the CTRW is decoupled, then the
probability density $P(x,t)$ of the walker position $X(t)$ depends
only on $w(x)$ and $p(\tau)$. According to \cite{MW}, in
Fourier-Laplace space this dependence has the form
\begin{equation}
    P_{ks} = \frac{1-p_{s}}{s(1-p_{s}w_{k})},
\label{M-Weq}
\end{equation}
where $w_{k} = \mathcal{F} \{w(x)\} = \int_{-\infty}^{\infty} dx
e^{ikx} w(x)$ ($-\infty <k <\infty$) is the Fourier transform of
$w(x)$, $p_{s} = \mathcal{L} \{p(t)\} = \int_{0}^{\infty} dt e^{-st}
p(t)$ $(\mathrm{Re} s>0)$ is the Laplace transform of $p( \tau)$, and
$P_{ks} = \mathcal{F} \{ \mathcal{L} \{P(x,t)\}\}$.

From Eq.~(\ref{M-Weq}) one can get
\begin{equation}
    P_{s}(x) = \frac{(1-p_{s})p_{s}}{s}\, \mathcal{F}^{-1}
    \bigg\{\frac{w_{k}}{1-p_{s}w_{k}} \bigg\} +
    \frac{1-p_{s}}{s}\, \delta(x)
\label{P(x)s}
\end{equation}
and
\begin{equation}
    P(x,t) = \mathcal{L}^{-1} \bigg\{\! \frac{(1-p_{s})
    p_{s}}{s}\mathcal{F}^{-1} \bigg\{\frac{w_{k}}{1-p_{s}
    w_{k}}\bigg\}\! \bigg\}\! + V(t) \delta(x).
\label{P(x,t)}
\end{equation}
Here, $\mathcal{F}^{-1} \{f_{k}\} = f(x) = (2\pi)^{-1} \int_{-
\infty}^{ \infty} dk e^{-ikx} f_{k}$ is the inverse Fourier
transform, $\delta(x)$ is the Dirac $\delta$ function, $\mathcal{L}
^{-1} \{g_{s}\} = g(t) = (2\pi i)^{-1} \int_{c- i\infty}^{c+ i\infty}
ds e^{st} g_{s}$ ($c$ is a real number exceeding the real parts of
all singularities of $g_{s}$) is the inverse Laplace transform, and
\begin{equation}
    V(t)= \mathcal{L}^{-1} \bigg\{ \frac{1-p_{s}}{s} \bigg\}
    = \int_{t}^{\infty} d\tau p(\tau)
\label{defV}
\end{equation}
with $V(0)=1$ and $V(\infty)=0$ is the survival or exceedance
probability. Using Eq.~(\ref{P(x,t)}), the integral formula
$\int_{-\infty}^{ \infty} dx e^{-ikx} = 2\pi\delta(k)$ and the
well-known properties of the $\delta$ function, it is not difficult
to show that the probability density $P(x,t)$ is properly normalized:
$\int_{- \infty} ^{\infty} dx P(x,t) =1$. Since $X(0) =0$, the
initial condition for $P(x,t)$ reads $P(x,0) = \delta(x)$ and, if
boundary conditions are not imposed, $P(x,t) \to 0$ as $t \to
\infty$.

According to this last property, the probability density of the
walker position vanishes in the long-time limit. It is therefore
reasonable to introduce the scaled walker position $Y(t) = a(t) X(t)$
and find the positive scaling function $a(t)$ such that the limiting
probability density
\begin{equation}
    \mathcal{P}(y) = \lim_{t \to \infty} \frac{1}{a(t)}
    \,P \! \left( \frac{y}{a(t)}, t \right)
\label{limP}
\end{equation}
of $Y(t)$, i.e., the probability density of the random variable
$Y(\infty)$, is non-vanishing and non-degenerate. The importance of
the functions $a(t)$ and $\mathcal{P}(y)$ is that, since $P(x,t) \sim
a(t) \mathcal{P}(a(t)x)$ as $t \to \infty$, they completely describe
the long-time behavior of the original walker position $X(t)$. To
satisfy the above requirements on $\mathcal{P}(y)$, the scaling
function must go to zero as $t \to \infty$ in a certain way. In fact,
these requirements permit one to determine $a(t)$ up to a constant
factor which, however, is not important and can be chosen for
convenience.

The pairs $a(t)$ and $\mathcal{P}(y)$ have been determined for all
cases characterized by finite second moments and/or heavy tails of
the probability densities $p(\tau)$ and $w(x)$ \cite{Kot}. In
contrast, the case with superheavy tails has been much less studied.
In fact, the pair $a(t)$ and $\mathcal{P}(y)$ has been determined
only when $p(\tau)$ has a superheavy tail and $w(x)$ has a finite
second moment $l_{2}$ \cite{DK2}. Because $l_{2} = \infty$ if $w(x)$
is heavy tailed, one may expect that in this case the long-time
behavior of the walker position changes qualitatively and thus the
pair $a(t)$ and $\mathcal{P}(y)$ changes as well. More precisely, in
this paper we study the long-time behavior of decoupled CTRWs whose
waiting-time densities $p(\tau)$ and jump densities $w(x)$ [it is
assumed that $w(-x) = w(x)$] are described by the asymptotic formulas
\begin{equation}
    p(\tau) \sim \frac{h(\tau)}{\tau} \quad (\tau \to \infty)
\label{p as}
\end{equation}
and
\begin{equation}
    w(x) \sim \frac{u}{|x|^{1+ \alpha} } \quad (|x| \to \infty),
\label{w as}
\end{equation}
where the positive function $h(\tau)$ varies slowly at infinity,
i.e., $h(\mu \tau) \sim h(\tau)$ as $\tau \to \infty$ for all
$\mu>0$, the tail index $\alpha$ is restricted to the interval
$(0,2]$, and $u>0$. The waiting-time and jump densities considered
here belong to the classes of superheavy- and heavy-tailed densities,
respectively. The difference between these classes consists in
different asymptotic behavior of the constituent probability
densities that, in turn, results in different properties of their
fractional moments. Specifically, while the fractional moments
$\int_{0}^{\infty} d\tau \tau^{\rho} p(\tau)$ of $p(\tau)$ are
infinite for all $\rho>0$, the fractional moments $\int_{-\infty}
^{\infty} dx |x|^{\rho} w(x)$ of $w(x)$ are infinite only if $\rho
\geq \alpha$. It should also be noted that the conditions $u>0$ and
$\alpha \in (0,2]$ are completely compatible with the normalization
condition $\int_{-\infty} ^{\infty} dx w(x) =1$. In contrast, the
normalization condition $\int_{0}^{\infty} d\tau p(\tau) =1$ imposes
an additional restriction on the asymptotic behavior of $h(\tau)$:
$h(\tau) = o(1/\ln \tau)$ as $\tau \to \infty$.

\section{SCALING FUNCTIONS AND THE LIMITING PROBABILITY DENSITY}
\label{Lim}

According to the Tauberian theorem for Laplace transforms \cite{Fel},
the long-time behavior of the probability density $P(x,t)$ is
determined by the asymptotic behavior of the Laplace transform
$P_{s}(x)$ when the \textit{real} parameter $s$ tends to zero.
Because the waiting-time distribution is normalized to unity, the
condition $p_{s} \to 1$ holds as $s \to 0$. It follows from
Eq.~(\ref{P(x)s}) that we also need to find the $s\to 0$ behavior of
$1-p_{s}$. To this end, it is convenient to use the representation
$1-p_{s} = \int_{0} ^{\infty} dq e^{-q} V(q/s)$ which, together with
the fact \cite{DK2} that the survival probability $V(t)$ varies
slowly at infinity, immediately gives
\begin{equation}
    1 - p_{s} \sim V(1/s)
\label{as ps}
\end{equation}
as $s \to 0$. Then, taking into account that as $s \to 0$ the main
contribution to $\mathcal{F}^{-1}\{w_{k}/ (1-p_{s}w_{k}) \}$ comes
from a small vicinity of the point $k=0$, i.e.,
\begin{equation}
    \mathcal{F}^{-1}\bigg\{\frac{w_{k}}{1-p_{s}w_{k}}
    \bigg\} \sim \mathcal{F}^{-1}\bigg\{\frac{1}
    {V(1/s) + 1-w_{k}} \bigg\},
\label{rel1}
\end{equation}
Eq.~(\ref{P(x)s}) in the small-$s$ limit yields
\begin{equation}
    P_{s}(x) \sim \frac{V(1/s)}{s} \delta(x) +
    \frac{V(1/s)}{\pi s}\int_{0}^{\infty}\! dk
    \frac{\cos(xk)}{V(1/s) + 1 - w_{k}}.
\label{as Ps(x)}
\end{equation}

The long-time behavior of $P(x,t)$ can be found directly from the
limiting formula (\ref{as Ps(x)}) by applying the above mentioned
Tauberian theorem. It states that if the function $v(t)$ is
ultimately monotonic and $v_{s} \sim s^{-\gamma} L(1/s)$ ($0<\gamma<
\infty$) as $s \to 0$, then $v(t) \sim t^{\gamma-1} L(t)/ \Gamma(
\gamma)$ as $t \to \infty$. Here, $\Gamma(\gamma)$ denotes the gamma
function and $L(t)$ is a slowly varying function at infinity. In our
case $\gamma =1$, therefore from Eq.~(\ref{as Ps(x)}) one obtains
\begin{equation}
    P(x,t) \sim V(t) \delta(x) + \frac{V(t)}{\pi}
    \int_{0}^{\infty} dk \frac{\cos(xk)}{V(t) + 1 - w_{k}}
\label{as P(x,t)}
\end{equation}
($t \to \infty$). Since in the long-time limit (when $V(t)$ tends to
zero) the main contribution to the integral in Eq.~(\ref{as P(x,t)})
comes from a small vicinity of the point $k=0$, the exact formula
\begin{equation}
    1 - w_{k} = 2\int_{0}^{\infty} dx [1-\cos (kx)] w(x)
\label{wk}
\end{equation}
can be replaced by one valid in this regime. Using Eq.~(\ref{w as})
and the integral relation
\begin{equation}
    \int_{0}^{\infty} dx \frac{1-\cos (x)}{x^{1+\alpha}}
    = \frac{\pi}{2\Gamma(1+\alpha) \sin(\pi \alpha/2)}
\label{rel2}
\end{equation}
($0<\alpha<2$), from Eq.~(\ref{wk}) at $|k| \to 0$ we find
\begin{equation}
    1 - w_{k} \sim \frac{\pi u}{\Gamma(1+\alpha)
    \sin(\pi \alpha/2)} |k|^{\alpha}.
\label{wk1}
\end{equation}

Substituting this result into the asymptotic formula (\ref{as
P(x,t)}) and applying the definition (\ref{limP}), the limiting
probability density $\mathcal{P}(y)$ can be written in the form
\begin{equation}
    \mathcal{P}(y) = \lim_{t \to \infty} \frac{1}{\pi}
    \int_{0}^{\infty}dx \frac{\cos (yx)}{1 +
    \frac{\pi u a^{\alpha}(t) }{\Gamma(1+\alpha)
    \sin(\pi \alpha/2)V(t)}\,x^{\alpha}}.
\label{P1}
\end{equation}
It appears from this that $\mathcal{P}(y)$ is non-vanishing and
non-degenerate only if the factor in front of $x^{\alpha}$ tends to a
nonzero finite limit as $t \to \infty$. Assuming for convenience that
this limit equals 1, we obtain the asymptotic representation of the
scaling function
\begin{equation}
    a(t) \sim \bigg( \frac{\Gamma(1 + \alpha)
    \sin (\pi \alpha/2)}{\pi u} V(t) \bigg)^{1/\alpha}
\label{a1}
\end{equation}
($t \to \infty$) and the corresponding limiting density
\begin{equation}
    \mathcal{P}(y) = \frac{1}{\pi} \int_{0}^{\infty}dx
    \frac{\cos (yx)}{1 + x^{\alpha}}
\label{P2}
\end{equation}
(the fact that $\mathcal{P}(y)$ is a probability density will be
proved below). The symmetry condition $\mathcal{P}(-y) = \mathcal{P}
(y)$, which follows from Eq.~(\ref{P2}), is a consequence of the
symmetry of the jump density $w(x)$.

Since at $\alpha=2$ the integral in Eq.~(\ref{rel2}) diverges, the
limiting formula (\ref{wk1}) is not applicable to this case.
Therefore, in order to find $1 - w_{k}$ at $\alpha = 2$ and $|k| \to
0$, we first split the interval of integration in Eq.~(\ref{wk}) into
two parts, $(0,b)$ and $(b, \infty)$ with $b \sim 1$. Then, taking
into account that as $|k| \to 0$ the contribution of the first
interval to the right-hand side of Eq.~(\ref{wk}) can be approximated
by $k^{2} \int_{0}^{b} dx x^{2}w(x)$ and the second one by $uk^{2}
\ln(1/|k|)$, we get
\begin{equation}
    1 - w_{k} \sim u k^{2} \ln \frac{1}{|k|}
\label{wk2}
\end{equation}
($|k| \to 0$). In accordance with this, the limiting probability
density when $\alpha = 2$ takes the form
\begin{equation}
    \mathcal{P}(y) = \lim_{t \to \infty} \frac{1}{\pi}
    \int_{0}^{\infty}dx \frac{\cos (yx)}{1 + \frac{ua^{2}(t)
    \ln [1/a(t)]}{V(t)}\, x^{2}}.
\label{P3}
\end{equation}

As before, we choose the long-time limit of the factor in front of
$x^{2}$ to be equal to unity. In this case the asymptotic behavior of
the scaling function $a(t)$ is determined by the relation $ua^{2}(t)
\ln [1/a(t)] \sim V(t)$ ($t \to \infty$). Assuming that $a(t) \sim
\sqrt{V(t)/u} \, a_{1}(t)$, where the positive function $a_{1}(t)$
satisfies the conditions $a_{1}(t) \to 0$ and $\sqrt{V(t)} =
o(a_{1}(t))$ as $t \to \infty$, from this relation we obtain
$a_{1}(t) \sim \sqrt{2/\ln [1/V(t)]}$, and thus
\begin{equation}
    a(t) \sim \sqrt{ \frac{2V(t)}{u \ln [1/V(t)] }}
\label{a2}
\end{equation}
($t \to \infty$). The limiting probability density (\ref{P3}) which
corresponds to this scale function is given by
\begin{equation}
    \mathcal{P}(y) = \frac{1}{\pi} \int_{0}^{\infty}dx
    \frac{\cos (yx)}{1 + x^{2}} = \frac{1}{2}e^{-|y|},
\label{P4}
\end{equation}
showing that Eq.~(\ref{P2}) is valid for $\alpha =2$ as well. We note
that the same two-sided exponential density (\ref{P4}) describes the
limiting distribution when the jump density $w(x)$ has a finite
second moment $l_{2} = \int_{- \infty}^{ \infty} dx x^{2}w(x)$
\cite{DK2}. However, because at $l_{2}<\infty$ the asymptotic
behavior of the scaling function, $a(t) \sim \sqrt{2V(t) /l_{2}}$, is
quite different from that given in Eq.~(\ref{a2}), the long-time
behaviors of the walker position in these cases are also quite
different.

\subsection{Positivity and unimodality of $\bm{\mathcal{P} (y)}$}
\label{posit}

To be a probability density, the function $\mathcal{P}(y)$ must be
normalized and positive (non-negative). The normalization condition
$\int_{-\infty} ^{\infty} dy \mathcal{P}(y) =1$ can easily be proved
using Eq.~(\ref{P2}), which represents $\mathcal{P}(y)$ as a cosine
Fourier transform, and the integral representation $\delta(x) =
(1/2\pi) \int_{-\infty}^{\infty} dy \cos (yx)$ of the $\delta$
function. However, except for the case $\alpha = 2$, where according
to Eq.~(\ref{P4}) $\mathcal{P}(y)>0$, the use of Eq.~(\ref{P2}) to
prove the positivity of $\mathcal{P}(y)$ is impractical because of
the oscillating character of the integrand. On this point, the
representation of $\mathcal{P}(y)$ in the form of a Laplace transform
would be preferable. In order to find it, we first define the
function
\begin{equation}
    f(z) = \frac{1}{\pi} \frac{e^{i|y|z}}{1 + z^{\alpha}}
\label{def_f}
\end{equation}
($0<\alpha<2$) of the complex variable $z=x+iu$. This function is
analytic in the first quadrant of the $z$-plane (when $|z|>0$ and
$0\leq \mathrm{arg}\,z \leq \pi/2$), and so from the Cauchy integral
theorem \cite{AF} we have $\oint_{C} dzf(z) =0$, where $C$ is a
simple closed contour that lies in the domain of analyticity of
$f(z)$. Then, choosing the contour $C$ to be the boundary of the
first quadrant (we emphasize that the branch point $z=0$ is outside
the contour) and applying the Jordan lemma \cite{AF}, the above
integral reduces to
\begin{equation}
    \int_{0}^{\infty}dx f(x) - i\int_{0}^{\infty}du f(iu) = 0.
\label{IntRel}
\end{equation}
Finally, taking into account that $\mathcal{P}(y) = \mathrm{Re}
\left[ \int_{0}^{\infty} dx f(x) \right]$ and $i^{\alpha} = \cos (\pi
\alpha/2) + i \sin (\pi \alpha/2)$, from the real part of
Eq.~(\ref{IntRel}) we obtain
\begin{equation}
    \mathcal{P}(y) = \frac{1}{\pi} \int_{0}^{\infty}dx
    e^{-|y|x}\frac{\sin (\pi \alpha/2) x^{\alpha}}{1 +
    2\cos (\pi \alpha/2) x^{\alpha} + x^{2\alpha}}.
\label{P2b}
\end{equation}

\begin{figure}
    \centering
    \includegraphics[totalheight=5cm]{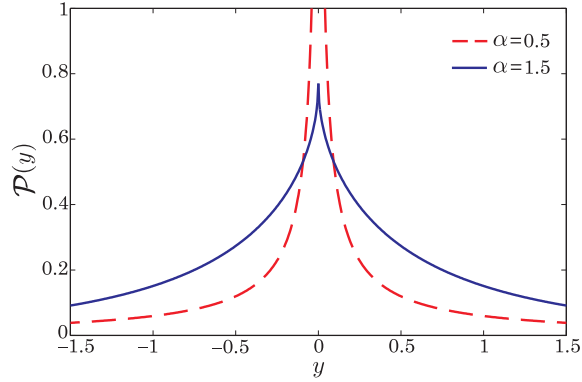}
    \caption{\label{fig1} Plots of the probability
    density $\mathcal{P}(y)$ for two values of the tail index
    $\alpha$ belonging to the intervals $(0,1]$ and $(1,2]$.}
\end{figure}

The main advantage of this representation of $\mathcal{P}(y)$ is that
it clearly shows that $\mathcal{P}(y)>0$ when $0<\alpha<2$. Thus,
since $\mathcal{P}(y)$ is positive for $\alpha=2$ as well, we can
conclude that the function $\mathcal{P}(y)$ is indeed the probability
density for all $\alpha$ in the interval $(0,2]$. Another important
property of $\mathcal{P}(y)$, which follows directly from
Eq.~(\ref{P2b}), is that $d\mathcal{P}(y)/dy <0$ when $y>0$. Together
with the condition $\mathcal{P}(-y) = \mathcal{P} (y)$, it shows that
the limiting probability density is symmetric, unimodal and centered
at the origin. In contrast to the scaling function, which depends on
all the parameters characterizing the asymptotic behavior of the
waiting time and jump densities, the limiting density depends only on
the tail index $\alpha$. According to Eqs.~(\ref{P2}) and
(\ref{P2b}), this parameter strongly influences the properties of
$\mathcal{P} (y)$. In particular, the behaviors of $\mathcal{P}(y)$
in the vicinity of the origin differ substantially from one another
when $\alpha \in (0,1]$ and $\alpha \in (1,2]$, as illustrated in
Fig.~\ref{fig1} (for details, see Sec.~\ref{Asymp}).

\section{LIMITING DISTRIBUTION IN TERMS OF SPECIAL FUNCTIONS}
\label{Rep}

To get more insight into the mathematical structure of the limiting
probability density $\mathcal{P}(y)$, it is reasonable to express it
in terms of well-known special functions. Toward this end, we first
represent $\mathcal{P}(y)$ as the inverse Mellin transform. The
Mellin transform of a function $f(y)$ is defined by $f_{r} =
\mathcal{M} \{ f(y) \} = \int_{0}^{ \infty} dy f(y) y^{r-1}$.
Therefore, for the function $f(y) = \int_{0}^{\infty} dx u(yx)v(x)$
one gets $f_{r} = u_{r}v_{1-r}$ \cite{DB}. If $f(y)$ is associated
with $\mathcal{ P}(y)$ from Eq.~(\ref{P2}), then the functions $u(x)$
and $v(x)$ can be chosen as $u(x) = \pi^{-1} \cos (x)$ and $v(x) = (1
+ x^{\alpha})^{-1}$ whose Mellin transforms are given by \cite{DB}
\begin{equation}
    u_{r} = \frac{1}{\pi} \Gamma(r) \cos\!
    \left( \frac{\pi r}{2} \right) \quad
    (0 < \mathrm{Re}\, r <1)
\label{ur}
\end{equation}
and
\begin{equation}
    v_{r} = \frac{1}{\alpha} \Gamma\! \left(
    \frac{r}{\alpha} \right)\! \Gamma\! \left(1-
    \frac{r}{\alpha} \right) \quad
    (0 < \mathrm{Re}\, r < \alpha).
\label{vr}
\end{equation}
Using the reflection formula \cite{Erd2} $\Gamma (1/2 - r/2) \Gamma
(1/2 + r/2) = \pi/ \cos(\pi r/2)$ to replace $\cos(\pi r/2)$ in
Eq.~(\ref{ur}), the Mellin transform $\mathcal{P}_{r} = u_{r}
v_{1-r}$ of $\mathcal{P}(y)$ takes the form
\begin{equation}
    \mathcal{P}_{r} =\frac{\Gamma (r) \Gamma (1-1/\alpha +
    r/\alpha) \Gamma(1/\alpha - r/\alpha)}
    {\alpha \Gamma (1/2-r/2) \Gamma (1/2+r/2)},
\label{Pr}
\end{equation}
where $\max {(1-\alpha,0)}<\mathrm{Re}\, r<1$. Finally, introducing
the inverse Mellin transform as $\mathcal{M}^{-1} \{ f_{r} \} = f(y)
= (2\pi i)^{-1} \int_{c- i\infty}^{c + i\infty} dr f_{r} y^{-r}$ and
utilizing the fact that $\mathcal{P}(-y) = \mathcal{P}(y)$, we find
\begin{equation}
    \mathcal{P}(y) = \frac{1}{2\pi i}
    \int_{c-i\infty}^{c+i\infty}dr \mathcal{P}_{r}|y|^{-r}.
\label{P5}
\end{equation}

The structure of $\mathcal{P}_{r}$ suggests that the probability
density $\mathcal{P}(y)$ is a particular case of the Fox $H$-function
which can be defined by means of a Mellin-Barnes integral as follows
(see, e.g., Ref.~\cite{MSH}):
\begin{equation}
    H_{p,q}^{m,n} \left[ y \Big |
    \begin{array}{lcl}
        (a_{1},A_{1}),\ldots,(a_{p},A_{p})\\
        (b_{1},B_{1}),\ldots,(b_{q},B_{q})
    \end{array}
    \right] = \frac{1}{2\pi i} \int_{L} dr
    \Theta_{r} y^{-r}.
\label{defH}
\end{equation}
Here,
\begin{equation}
    \Theta_{r} =\frac{\prod_{j=1}^{m}\Gamma(b_{j} + B_{j}r)
    \prod_{j=1}^{n}\Gamma(1 - a_{j} - A_{j}r)}
    {\prod_{j=m+1}^{q}\Gamma(1 - b_{j} - B_{j}r)
    \prod_{j=n+1}^{p}\Gamma(a_{j} + A_{j}r)},
\label{Theta}
\end{equation}
$m,n,p,q$ are whole numbers, $0 \leq m \leq q$, $0 \leq n \leq p$,
$a_{j}$ and $b_{j}$ are real or complex numbers, $A_{j}, B_{j} >0$,
and $L$ is a suitable contour in the complex $r$-plane which
separates the poles of the gamma functions $\Gamma(b_{j} + B_{j}r)$
from the poles of the gamma functions $\Gamma(1 - a_{j} - A_{j}r)$.
It is also assumed that an empty product equals 1. Comparing
Eqs.~(\ref{Pr}) and (\ref{P5}) with Eqs.~(\ref{Theta}) and
(\ref{defH}), respectively, we see that
\begin{equation}
    \mathcal{P}(y)=\frac{1}{\alpha}H_{2,3}^{2,1}
    \left[ |y| \Big |
        \begin{array}{lcl}
        (1-1/\alpha,1/\alpha), (1/2,1/2)\\
        (0,1),(1-1/\alpha,1/\alpha),(1/2,1/2)
        \end{array}
    \right].
\label{FoxP}
\end{equation}

It should be noted that the cumulative distribution function $F(y) =
1/2 +\int_{0}^{y} dy' \mathcal{P}(y')$ of the random variable
$Y(\infty)$ can also be expressed through the $H$-function. To show
this, we write  $\widetilde F(y)=\int_{0}^{y} dy' \mathcal{P}(y')$
and take into account the following property of the Mellin transform
\cite{DB}: $\mathcal{M}\left\{\int_0^y f(y') dy'\right\}=-f_{r+1}/r$.
According to this, $ \widetilde F_r=-\mathcal{P}_{r+1}/r$ and, from
Eq.~\eqref{Pr} and the functional equation $\Gamma(1+x) =
x\Gamma(x)$, one gets
\begin{equation}
    \widetilde F_r = -\frac{\Gamma (r) \Gamma (r/\alpha)
    \Gamma (1 - r/\alpha) }{\alpha \Gamma (1 - r/2)
    \Gamma (r/2)}
\label{Gr}
\end{equation}
[$-\min {(1,\alpha)}<\mathrm{Re}\, r<0$]. Therefore, using
Eqs.~(\ref{defH}) and (\ref{Theta}), we obtain
\begin{equation}
    F(y)= \frac{1}{2} -\frac{\mathrm{sgn} (y)}
    {\alpha}H_{2,3}^{2,1} \left[ |y| \Big |
    \begin{array}{lcl}
        (0,1/\alpha),(0,1/2)\\
        (0,1),(0,1/\alpha),(0,1/2)
    \end{array}
    \right].
\label{F3}
\end{equation}

\subsection{Particular examples}
\label{Part}

For some special values of the tail parameter $\alpha$ the
$H$-functions in Eqs.~(\ref{FoxP}) and (\ref{F3}) can be reduced to
more familiar special (or even elementary) functions. Because the
probability density $\mathcal{P}(y)$ and the distribution function
$F(y)$ provide equivalent descriptions of the long-time behavior of
the scaled walker position $Y(t)$, next we consider only the
properties of $\mathcal{P}(y)$. The simplest situation occurs when
$\alpha=2$. In this case both reduction formulas \cite{MSH} can be
applied, yielding
\begin{eqnarray}
    \mathcal{P}(y) \!\!&=&\!\! \frac{1}{2}H_{2,3}^{2,1}
    \left[ |y| \Big |
        \begin{array}{lcl}
        (1/2,1/2), (1/2,1/2)\\
        (0,1),(1/2,1/2),(1/2,1/2)
        \end{array}
    \right]
    \nonumber\\ [4pt]
    \!\!&=&\!\!\frac{1}{2}H_{1,2}^{2,0}
    \left[ |y| \Big |
    \begin{array}{lcl}
        (1/2,1/2)\\
        (0,1),(1/2,1/2)
    \end{array}
    \right]
    \nonumber\\ [4pt]
    \!\!&=&\!\!\frac{1}{2}H_{0,1}^{1,0}
    \left[ |y| \Big |
    \begin{array}{lcl}
        \\
        (0,1)
    \end{array}
    \right].
\label{P_2}
\end{eqnarray}
Since the last $H$-function equals $e^{-|y|}$ \cite{MSH}, this
ascertains that Eq.~(\ref{FoxP}) at $\alpha = 2$ reduces to
Eq.~(\ref{P4}).

If the parameter $\alpha$ is rational, then the probability density
$\mathcal{P}(y)$ can, in principle, be expressed in terms of the
Meijer $G$-function as well. The $G$-function, which is a particular
case of the $H$-function, is defined as
\begin{equation}
    G_{p,q}^{m,n} \left[ y \Big |
    \begin{array}{lcl}
        a_{1},\ldots,a_{p}\\
        b_{1},\ldots,b_{q}
    \end{array}
    \right] = \frac{1}{2\pi i} \int_{L} dr
    \Psi_{r} y^{-r}
\label{defG}
\end{equation}
with $\Psi_{r} = \Theta_{r}|_{A_{j}, B_{j} = 1}$. As a first
illustrative example, we consider the case when $\alpha =1$. Changing
the variable of integration in Eq.~(\ref{P5}) from $r$ to $2r$, one
readily obtains $\mathcal{P}(y) =(\pi i)^{-1} \int_{c -i\infty} ^{c
+i\infty}dr \mathcal{P}_{2r}(y^{2})^{-r}$, where $\max {(1-\alpha,0)
}/2<c<1/2$. Then, using Eq.~(\ref{Pr}) with $\alpha=1$ and the
duplication formula \cite{Erd2} $\Gamma (2r) = \pi^{-1/2} 2^{2r-1}
\Gamma(r) \Gamma(1/2 +r)$, the Mellin transform $\mathcal{P} _{2r}$
can be written in the form
\begin{equation}
    \mathcal{P}_{2r} = \frac{2^{2r}}{4\pi^{3/2}}
    \Gamma^{2}(r)\Gamma (1/2 + r) \Gamma(1 - r).
\label{P2r1}
\end{equation}
Therefore, in accordance with the definition (\ref{defG}), the
limiting probability density (\ref{FoxP}) at $\alpha=1$ has the
following $G$-function representation:
\begin{eqnarray}
    \mathcal{P}(y) \!\!&=&\!\! H_{2,3}^{2,1}
    \left[ |y| \Big |
        \begin{array}{lcl}
        (0,1), (1/2,1/2)\\
        (0,1),(0,1),(1/2,1/2)
        \end{array}
    \right]
    \nonumber\\ [4pt]
    \!\!&=&\!\!\frac{1}{2\pi ^{3/2}}G_{1,3}^{3,1}
    \left[ y^2/4 \Big |
    \begin{array}{lcl}
        0\\
        0,0,1/2
    \end{array}
    \right].
\label{P_1a}
\end{eqnarray}
Remarkably, the limiting probability density $\mathcal{P}(y)$ at
$\alpha =1$ can be expressed not only in terms of the Fox and Meijer
functions, but also in terms of the well-known sine, $\mathrm{si}(y)
= -\int_{y}^{\infty} dx \sin (x)/x$, and cosine, $\mathrm{Ci}(y) =
-\int_{y}^{\infty} dx \cos (x)/x$, integral functions. Indeed, using
the exact result for the cosine Fourier transform of $(1+x)^{-1}$
\cite{Erd1}, we obtain
\begin{equation}
    \mathcal{P}(y) = -\frac{1}{\pi} [\sin(|y|)\,
    \mathrm{si}(|y|) + \cos(y)\, \mathrm{Ci}(|y|)].
\label{P 1b}
\end{equation}

Finally, in our last example we consider the case $\alpha =1/2$.
Following straightforward calculations similar to those described
above, for the Mellin transform $\mathcal{P}_{2r}$ we obtain the
expression
\begin{eqnarray}
    \mathcal{P}_{2r} \!\!&=&\!\! \frac{2^{2r}}
    {8\pi^{7/2}} \Gamma(-1/4 +r)\Gamma^{2}(r)
    \Gamma(1/4 +r)\Gamma (1/2 + r)
    \nonumber\\ [4pt]
    \!\!&&\!\! \times \Gamma(5/4 -r)
    \Gamma(1 - r) \Gamma(3/4 -r)
\label{P2r2}
\end{eqnarray}
from which it follows that
\begin{eqnarray}
    \mathcal{P}(y) \!\!&=&\!\! 2H_{2,3}^{2,1}
    \left[ |y| \Big |
        \begin{array}{lcl}
        (-1,2), (1/2,1/2)\\
        (0,1),(-1,2),(1/2,1/2)
        \end{array}
    \right]
    \nonumber\\ [4pt]
    \!\!&=&\!\!\frac{1}{4\pi ^{7/2}}G_{3,5}^{5,3}
    \left[ y^2/4 \Big |
    \begin{array}{lcl}
        -1/4,0,1/4\\
        -1/4,0,0,1/4,1/2
    \end{array}
    \right]. \qquad
\label{P_1/2}
\end{eqnarray}

\section{ASYMPTOTIC BEHAVIOR OF $\bm{\mathcal{P} (y)}$}
\label{Asymp}

Using Eq.~(\ref{FoxP}), the behavior of the limiting probability
density $\mathcal{P}(y)$ for small and large values of $|y|$ can, in
principle, be found from the expansions obtained for the $H$-function
in different limits (for details, see Ref.~\cite{MSH} and references
therein). However, because $\mathcal{P}(y)$ is a very particular case
of the $H$-function, it is reasonable and convenient to derive the
corresponding limiting formulas directly from the source
representation (\ref{P2}).

\subsection{Short-distance behavior}
\label{Short}

There are three regions of the tail index $\alpha$, which we consider
separately, where the limiting behaviors of $\mathcal{P}(y)$ as $|y|
\to 0$ differ from one another.

$\mathit{\alpha \in (0,1)}$. In this case Eq.~(\ref{P2}), after
changing the variable of integration from $x$ to $x/|y|$, as $|y| \to
0$ yields
\begin{equation}
    \mathcal{P}(y) \sim \frac{1}{\pi |y|^{1-\alpha}}
    \int_{0}^{\infty} dx \frac{\cos (x)}{x^{\alpha}}.
\label{asP1}
\end{equation}
Then, since $\int_{0}^{\infty} dx \cos (x)/x^{\alpha} = \Gamma(1 -
\alpha) \sin(\pi \alpha/2)$, one gets
\begin{equation}
    \mathcal{P}(y) \sim \frac{\Gamma(1 - \alpha)
    \sin(\pi \alpha/2)}{\pi |y|^{1-\alpha}}.
\label{asP2}
\end{equation}

$\mathit{\alpha = 1}$. Using the formulas $\mathrm{si} (|y|) \sim
|y|$ and $\mathrm{Ci}(|y|) \sim \ln{|y|}$ ($|y| \to 0$) \cite{AS},
Eq.~(\ref{P 1b}) which follows from Eq.~(\ref{P2}) immediately yields
\begin{equation}
    \mathcal{P}(y) \sim -\frac{1}{\pi } \ln |y|.
\label{asP3}
\end{equation}

$\mathit{\alpha \in (1,2]}$. Finally, in this case it is convenient
to rewrite Eq.~(\ref{P2}) in the form
\begin{equation}
    \mathcal{P}(y) = \mathcal{P}(0) - \frac{|y|^{\alpha -1}}
    {\pi} \int_{0}^{\infty}dx \frac{1-\cos (x)}
    {|y|^{\alpha} + x^{\alpha}},
\label{P6}
\end{equation}
where $\mathcal{P}(0) = [\alpha \sin(\pi/\alpha)]^{-1}$. Then,
neglecting $|y|^{\alpha}$ in the integrand and taking into account
that $\int_{0}^{\infty}dx [1-\cos (x)]/x^{ \alpha} = \Gamma(2-
\alpha) \sin(\pi \alpha/2)/(\alpha -1)$, we obtain
\begin{equation}
    \mathcal{P}(y) \sim \frac{1}{\alpha \sin (\pi/\alpha)}
    - \frac{\Gamma(2 - \alpha) \sin(\pi \alpha/2)}
    {\pi (\alpha -1)}|y|^{\alpha -1}.
\label{asP4}
\end{equation}
It should be noted that, since $\lim_{x \to 0} \Gamma(x) \sin(\pi
x/2) = \pi/2$, the limiting formula (\ref{asP4}) at $\alpha=2$
reduces to $\mathcal{P}(y) \sim (1 - |y|)/2$, in accordance with
Eq.~(\ref{P4}).

\subsection{Long-distance behavior}
\label{Long}

The asymptotic behavior of $\mathcal{P}(y)$ as $|y| \to \infty$ can
easily be found by a single (if $0<\alpha<1$) or double (if
$1<\alpha<2$) integration by parts of Eq.~(\ref{P2}) with a
subsequent change of the integration variable from $x$ to $x/|y|$. In
particular, for $\alpha \in (0,1)$ this yields
\begin{eqnarray}
    \mathcal{P}(y) \!\!&=&\!\! \frac{\alpha}{\pi |y|^{1+\alpha}}
    \int_{0}^{\infty}dx \frac{\sin (x)}
    {x^{1-\alpha}[1 + (x/|y|)^{\alpha}]}
    \nonumber\\ [4pt]
    \!\!&\sim&\!\! \frac{\alpha}{\pi |y|^{1+\alpha}}
    \int_{0}^{\infty}dx \frac{\sin (x)}{x^{1-\alpha}}
\label{rel3}
\end{eqnarray}
and so
\begin{equation}
    \mathcal{P}(y) \sim \frac{\Gamma(1 + \alpha)
    \sin(\pi \alpha/2)}{\pi |y|^{1+\alpha}}.
\label{asP5}
\end{equation}
It is not difficult to verify that the asymptotic formula
(\ref{asP5}) also holds for $\alpha \in (1,2)$. Moreover, since
Eq.~(\ref{P 1b}) leads to $\mathcal{P}(y) \sim \pi^{-1} |y|^{-2}$ as
$|y| \to \infty$, this formula is valid for $\alpha=1$ as well.

Thus, according to Eq.~(\ref{asP5}), the limiting probability density
$\mathcal{P}(y)$ when $\alpha \in (0,2)$ is heavy-tailed with the
same tail index $\alpha$ as in the jump density $w(x)$. In contrast,
at $\alpha = 2$ the limiting density has exponential tails, while the
jump density is still heavy-tailed, see Eq.~(\ref{w as}). We also
note that the same tail index $\alpha$ characterizes the limiting
probability density when both the waiting-time and jump distributions
are heavy-tailed \cite{Kot}. However, this does not mean that the
long-time behaviors of the CTRWs with heavy- and superheavy-tailed
distributions of waiting times are identical. This is because the
scaling functions for these CTRWs are quite different. Specifically,
while in the former case the scaling functions are power functions of
time \cite{Kot}, in the latter case they vary more slowly, see
Eqs.~(\ref{a1}) and (\ref{a2}).

\section{CONCLUSIONS}
\label{Concl}

We have determined a new class of asymptotic solutions of the CTRWs
characterized by superheavy-tailed distributions of waiting times and
symmetric heavy-tailed distributions of jump lengths. These solutions
represent the probability densities of the scaled walker position,
i.e., the random walker position multiplied by a time-dependent
deterministic scaling function, in the long-time limit. We have found
both the limiting probability densities and the corresponding scaling
functions which completely describe the long-time behavior of the
reference CTRWs. It turns out that the scaling functions depend on
the survival probability characterizing the long-time behavior of the
waiting-time density and on the tail index $\alpha \in (0,2]$
describing the asymptotic behavior of the jump density. In contrast,
the limiting densities, which have been represented in the form of
Fourier and Laplace transforms, depend only on $\alpha$.

The limiting probability densities $\mathcal{P} (y)$ form a class of
symmetric and unimodal functions centered at the origin. Among other
things, we have determined the limiting behavior of these densities
for small and large distances. We find that while at $\alpha =2$ the
function $\mathcal{P} (y)$ has exponential tails, at $\alpha \in
(0,2)$ the tails are heavy and are characterized by the same tail
index $\alpha$ as the jump density. In the vicinity of the origin,
the behavior of $\mathcal{P} (y)$ for $\alpha \in (0,1]$ is quite
different from that for $\alpha \in (1,2]$. Specifically,
$\mathcal{P} (0)$ is infinite in the former case and is finite in the
latter. Finally, we have expressed the limiting probability densities
in terms of the Fox $H$-function for the general case of arbitrary
$\alpha$ and, for a few values of $\alpha$, in terms of the Meijer
$G$-function.

\section*{ACKNOWLEDGMENTS}

S.I.D.~is grateful to the Ministry of Education and Science of
Ukraine for the financial support and the Max-Planck-Institut f\"{u}r
Physik komplexer Systeme, Dresden, for the hospitality during his
visit. S.B.Y.~acknowledges the financial support of the Ministerio de
Ciencia e Innovaci\'{o}n (Spain) through Grant No.~FIS2010-16587
(partially financed by FEDER funds) and of the Junta de Extremadura
through Grant No.~GR10158. K.L.~gratefully acknowledges the US
National Science Foundation under Grant No.~PHY-0855471.

\end{document}